# Hadron Properties just before Deconfinement.

G. Boyd[1,3], Sourendu Gupta[2,3], F. Karsch[1,3],
E. Laermann[1], B. Petersson[1] and K. Redlich[1,4]

[1] Fak. f. Physik, Univ. Bielefeld, Postfach 100131, D-33501 Bielefeld, Germany
[2] Present Address: TIFR, Homi Bhabha Road, Bombay 400005, India
[3] HLRZ, c/o Forschungszentrum Jülich, D-52425 Jülich, Germany
[4] Institute for Theoretical Physics, University of Wroclaw, PL-50205, Wroclaw, Poland

## Abstract

We have investigated hadron screening masses, the chiral condensate, and the pion decay constant close to the deconfinement phase transition in the confined phase of QCD. The simulations were done in the quenched approximation, on a lattice of size $32^3 \times 8$. We examined temperatures ranging from $0.75 T_c$ up to $0.92 T_c$. We see no sign of a temperature dependence in the chiral condensate or the meson properties, but some temperature dependence for the nucleon screening mass is not excluded.

# 1 Introduction

Despite extensive studies of the QCD phase transition [1–3] and the phases on either side of it, little is known with certainty about the changes in the excitation spectrum. In particular, the properties of the (quasi)-particle spectrum in the confined phase close to $T_c$ require a more thorough understanding, as the temperature dependence of hadron masses and other hadronic parameters will lead to observable consequences in current, and especially forthcoming, heavy ion collisions. For example, a temperature dependence of the $\rho$ mass and width leads to large modifications of the dilepton spectrum [4,5].

The temperature dependence of various hadronic properties has been addressed using a number of different approaches. These calculations yield different predictions for the behaviour of some quantities. For example, in chiral perturbation theory it has been found that the chiral condensate and $f_\pi$ decrease, while $m_\pi$ and $m_N$ increase [6] with increasing temperature. In an approach which uses a Skyrme-type Lagrangian, a relationship between various hadronic quantities was obtained [7] which suggests that all quantities should behave in the same way with increasing temperature. On the other hand, QCD sum rule predictions (eg. [5,8,9]) yield a decreasing nucleon mass and only a slow variation of $m_\rho$. On the other hand, it has been argued [10] that hadron masses calculated from finite temperature sum rules are temperature independent up to $O(T^2)$.

Since the above approaches have limited applicability at high temperatures, one may hope that a lattice calculation of the variation with temperature of any of these quantities below the phase transition may shed some light on these discrepancies.

At finite temperature the notion of the mass is somewhat ambiguous. One has to distinguish the *pole mass* defined through the pole position of the propagator from the *screening mass* obtained from the hadronic correlation functions at large spatial separations [11–13]. These two will coincide only if the ordinary zero temperature dispersion relations also hold at non-zero temperature. Most of the lattice calculations at finite temperature which have determined screening masses [12–20], concentrated on the region $T > T_c$. The results are compatible with the propagation of deconfined quarks. The few calculations performed for $T < T_c$ were, however, all at rather small temperatures, and no significant changes from the $T = 0$ masses were seen [21].

In this paper we present results from a study of hadronic properties in quenched QCD at temperatures between $0.75T_c$ and $0.92T_c$. We will discuss the temperature dependence of the pion, rho and nucleon screening masses, the chiral condensate, and the pion decay constant. Details of the simulations performed here are presented in section 2. Section 3 contains the results pertaining to the chiral sector, and section 4 those for the nucleon and $\rho$-meson. Finally, in section 5 we present our conclusions.

# 2 Simulations

Previous studies have shown that the quenched approximation yields, for both the zero temperature masses and the high temperature screening masses, results that are within 10% of those obtained using full QCD with both two and four fermion flavours. Fur-thermore, one sees a rapid, simultaneous change in the Polyakov loop and the chiral condensate in both the quenched approximation and in full QCD. This suggests that the hadronic properties are largely dependent on the gluon sector. We have thus chosen to use the quenched approximation for this study.

We have to point out however, that close to the critical temperature the fluctuations associated with a second order phase transition may have a large effect on the temperature dependence of the screening masses. In this region the two flavour case may thus yield different results to both the four flavour and the quenched cases. The studies of the chiral transition in two flavour QCD [22] suggest, however, that these effects only become relevant very close to $T_c$. Hence we expect that our present analysis is not yet affected by these effects.

We used a lattice of size $32^3 \times 8$ at coupling constants of $\beta = 5.90, 5.95, 5.975$ and $6.0$, giving a range in temperature from $0.75 T_c$ to $0.92 T_c$. We have set the temperature scale using the calculation of the non-perturbative beta function by the TARO collaboration [23], which indicates that the critical coupling lies at $\beta \sim 6.06$. This value is somewhat larger than those quoted in [24–26]. It is, however, supported by a direct determination of the critical point on $32^3 \times 8$ lattices, which is currently in progress [27].

Studies of the temperature dependence of hadronic properties close to $T_c$ need large spatial volumes in order to prevent flips between coexisting phases. The volume used here was seen to be large enough to prevent flips to the deconfined phase, even at $\beta = 6.00$.

We used four staggered valence quarks with masses $m_q = 0.05, 0.025$ and $0.01$ for the point sources, and a mass of $m_q = 0.01$ for the wall sources. It is known that quenching introduces chiral logarithms which may spoil the chiral limit. However, they do not seem to present a difficulty for the quantities and within the parameter range studied here [28].

Details of the configurations generated are shown in Table 1. The first set of configurations was generated using a Kennedy-Pendleton heat bath algorithm (indicated by HB), and the second set using a combination of heat bath and over-relaxed algorithms (indicated by OR). For the pure heat bath algorithm we used 50 sweeps per configuration, whilst for the combined algorithm we used four over-relaxed sweeps per heat bath, and 40 heat bath sweeps per configuration.

| $\beta$ | $T/T_c$ | No. Confs | $m_q$ |
| --- | --- | --- | --- |
| 5.90 | 0.75 | 150 (HB) | 0.05, 0.035, 0.025, 0.01 |
| 5.95 | 0.85 | 150 (HB) | 0.05, 0.025, 0.01 |
| 5.975 | 0.90 | 150 (HB) + 100 (OR) | 0.05, 0.025, 0.01 |
| 6.0 | 0.92 | 350 (OR) | 0.05, 0.025, 0.01 |

Table 1: The number of quenched configurations generated at each coupling is listed above, along with the valence quark masses used. The algorithm used was either pure heat-bath (HB) or a combined heat-bath and over-relaxation algorithm (OR).

The correlators obtained for each channel have been fitted using both the full correlation matrix as well as only the diagonal elements. Errors have been calculated using both the jackknife procedure and the standard $\chi^2 + 1$ definition. We found that in most cases both fitting procedures yield almost identical values for the fitted parameters and the jackknife errors, in particular when the errors on the propagators were small or when the fit interval could be limited to large distances from the source. Based on this, and the findings of [29], we have quoted masses based on fits using the diagonal elements of the correlation matrix for those cases where the fit with the full matrix did not work. The errors reflect the error from jackknife estimates, plus an estimate of the additional error coming from varying the fitting range.

## 3  Chiral Sector

Within the context of QCD sum rules as well as chiral perturbation theory the variation of hadronic parameters is closely related to the temperature dependence of fermionic and gluonic condensates. We thus will first discuss an analysis of the temperature dependence of the chiral condensate. It is known to change drastically at the deconfinement transition point in quenched QCD. However, a detailed investigation of its temperature dependence below $T_c$ on large lattices did not exist so far.

We have calculated the chiral condensate directly from the trace of the fermion matrix ($\langle\bar\psi\psi\rangle_{\text{SE}}$) and from the pseudoscalar (PS) and scalar (SC) propagators, which corrects for the linear dependence on the quark mass [30]. The latter method is given by the formula

$$\langle\bar\psi\psi\rangle_{\text{Int}} := \left(1 - m_{\text{q}}\frac{\partial}{\partial m_{\text{q}}}\right)\langle\bar\psi\psi\rangle = m_{\text{q}}\Big(\tilde G_{PS}(p=0) - \tilde G_{SC}(p=0)\Big). \qquad (3.1)$$

The meson propagators on the right hand side are taken at zero four momentum.

Figure 1(a) shows the value of the chiral condensate as a function of bare quark mass at a coupling $\beta = 6.0$, corresponding to $0.92T_c$. One can see quite clearly in the figure that both methods extrapolate to the same value at zero quark mass. The same holds true at the other couplings studied.

The temperature dependence of the chiral condensate is shown in figure 1(b), where we plot the ratio of the finite temperature chiral condensate extrapolated to zero quark mass with that at $T = 0$. For comparison results for two [22], three [31–33] and four [34, 35] flavour QCD, and another quenched study [21] have been included in the figure. These additional values have not been extrapolated to zero quark mass, which may lead to an underestimate of the temperature effects. However, it is clear from the figure that the chiral condensate does not change significantly until one is actually at the critical temperature. Even at the largest beta value, corresponding to $T = 0.92T_c$, the drop in the condensate is not statistically significant.

A second quantity of considerable interest for understanding the temperature dependence of the chiral sector of QCD is the pion decay constant. It can be determined directly from the relevant matrix element [30]

$$\sqrt{2}f_\pi m_\pi = -\langle 0|\bar u\gamma_4\gamma_5 d|\pi^+(\mathbf{p}=0)\rangle \quad. \qquad (3.2)$$

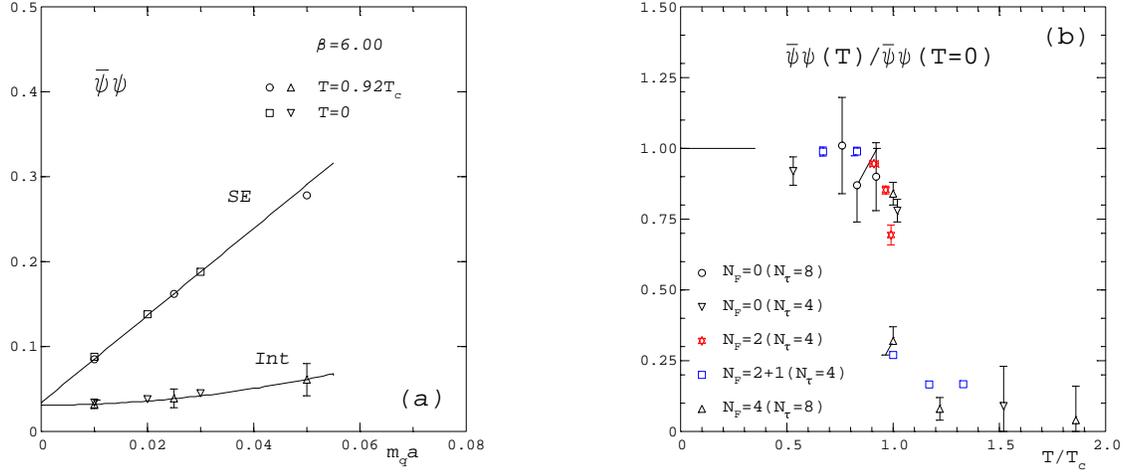

Figure 1: Figure (a) shows the chiral condensate as a function of the bare quark mass, at $\beta = 6.00$. The upper curve is obtained using the value taken directly from the trace of the fermion matrix, the lower curve after subtracting the linear dependence on the quark mass. The $T = 0$ results have been taken from [37]. Figure (b) shows the finite temperature chiral condensate normalised to its value at zero temperature. The circles represent our values, the inverted triangles a quenched calculation in [21] at $N_t = 4$, both extrapolated to vanishing quark mass and normalised to the data of [37,38]. The triangles show the result using the intercept method for a four flavour calculation [34,35] at $N_t = 8$, squares a 2 + 1 flavour calculation at $m_q = 0.0125$ in [31–33] at $N_t = 4$ and stars the results for 2 flavours at $m_q = 0.02$. [22].

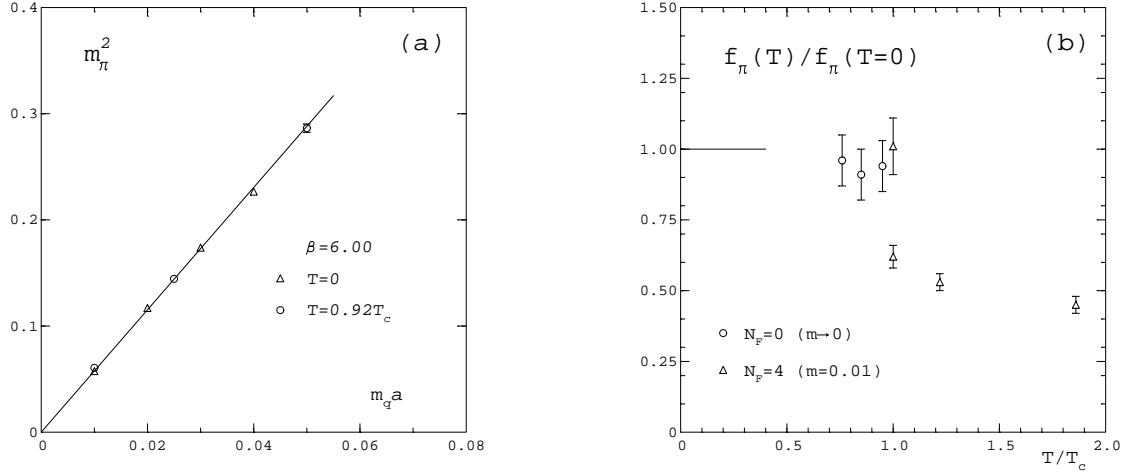

Figure 2: Figure (a) shows the square of the pion screening mass at $\beta = 6.00$ plotted as a function of the bare quark mass. The circles represent our data, the triangles the data of [36] at zero temperature. Figure (b) shows $f_\pi$ at $T > 0$ normalised to the value at $T = 0$ [37], at a quark mass of $m_q = 0.01$ in lattice units. For reference we include the values obtained for the matrix element above the phase transition as well.

At zero temperature $f_\pi$ is related to the chiral condensate and the pion mass through the Gell-Mann, Oakes, Renner (GMOR) relation,

$$m_\pi^2 f_\pi^2 = m_q \langle \bar{\psi}\psi \rangle_{(m_q=0)} \quad . \tag{3.3}$$

This relation is expected to be valid as long as chiral symmetry remains spontaneously broken.

A calculation of $f_\pi$ and a test of the GMOR relation at finite temperature requires a calculation of the pion mass. We have calculated at four values of the temperature the quark mass dependence of the pion screening mass. In all cases we find the expected behaviour for a Goldstone particle, $m_\pi^2 = A_\pi m_q$. The amplitude, $A_\pi$, does not show any temperature dependence. In figure 2(a) we have plotted the pion screening mass squared as a function of quark mass at a coupling of $\beta = 6.0$, along with the results at zero temperature. It is clear that both sets of points lie on the same straight line. There is no dependence on temperature; the relation $m_\pi^2 \propto m_q$ holds true, up to $T = 0.92T_c$.

The temperature independence of the amplitude, $A_\pi$, together with our finding of a temperature independent chiral condensate up to $T = 0.92T_c$ indicates that the pion decay constant will also be temperature independent up to this temperature, if the GMOR relates still holds. Indeed we have found, at all couplings, that the pion decay constant determined from eqn. 3.2 and extrapolated to zero quark mass agrees very well with that extracted from the GMOR relation using the value of the chiral condensate extrapolated to zero quark mass. Thus the GMOR is indeed valid at temperatures as high as $T = 0.92T_c$.

In figure 2(b) we have plotted the value of the pion decay constant at the lightest quark mass used, $m_q = 0.01$, normalised to the value at zero temperature as a function of $T/T_c$. As expected we see no temperature dependence up to $T = 0.92T_c$.

## 4   Nucleon and meson masses

The calculation of hadron masses usually proceeds through more or less involved fits to the hadron correlation functions $C_H(z)$ and requires a discussion of the fitting range etc. Another more direct way to discuss the long distance behaviour of the correlation functions is the analysis of local screening masses obtained from ratios of the hadron correlation functions at neighbouring lattice sites.

These are obtained by solving the following equation for the mass $m$:

$$\frac{C_H(z)}{C_H(z+1)} = \frac{\cosh(m(z - N_\sigma/2))}{\cosh(m(z+1 - N_\sigma/2))}. \tag{4.1}$$

For large distances these ratios stabilize and yield the lowest energy in the quantum number channel selected by the correlation function under consideration. As we are considering correlation functions at large spatial separation, the relation of these energies to screening masses differs at finite temperature for bosonic and fermionic states. The latter also receive a contribution from the non-vanishing Matsubara energy, $p_0 = \pi T$, i.e.

$$E_H^2 = m_H^2 + k \sin^2(\pi/N_\tau). \tag{4.2}$$

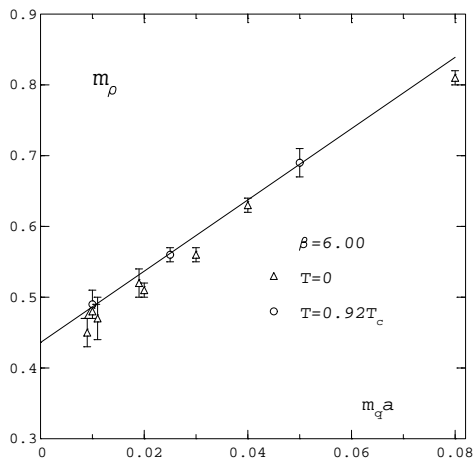

Figure 3: The rho screening mass at $\beta = 6.00$ plotted as a function of the bare quark mass. The circles represent our results using point sources, except for $m_q = 0.01$ where the value obtained from wall sources is given. The triangles represent the zero temperature results from [36, 37, 39, 40].

The rho screening mass at $T = 0.92T_c$ is shown in figure 3 as a function of the quark mass, including the zero temperature values. The local screening masses as a function of distance, as well as lines indicating the value obtained from a fit to the full propagator here and at zero temperature are shown in figure 4(a). It is clear from both figures that the rho screening mass does not show any significant dependence on temperature. As can be read off Table 2, where we have summarized our results for the screening masses, the rho screening mass is, within errors, identical in both pseudo-vector and vector-tensor channels, indicating that the continuum flavour symmetry is restored in the vector meson spectrum. The table also includes our numbers for the scalar, the $f_0$, where as usual the quarkline disconnected contribution has been omitted. Note that independent of the temperature, the $f_0$ exhibits a constant ratio to the $\rho$ mass, $m_{f_0}/m_\rho \simeq 1.2(1)$.

Finally, we turn to the nucleon. This was examined using wall sources at a quark mass of 0.01 in lattice units. The local screening masses for $T = 0.92T_c$, and the screening mass obtained from a full fit, are shown in figure 4(b) along with the result obtained at zero temperature. The increase in temperature clearly has a dramatic effect on the nucleon. However, most of this can be understood purely in terms of the fact that the lowest momentum of the nucleon is not zero, but $\pi T$, as discussed above.

The screening mass at $T = 0.92T_c$, extracted using eqn. 4.2 with the assumption that $k = 1$, indicates that the mass rises slightly: $m_N(0.92T_c) = (1.1 \pm 0.03)m_N(T = 0)$. We note, however, that there is also the possibility of a modification of the energy dispersion relation at finite temperature, which may lead to deviation of $k$ from unity [41]. The error given above is from the statistical error alone, as we cannot determine the systematic error which may be introduced by our assuming that eq. 4.2 with $k = 1$ is applicable.

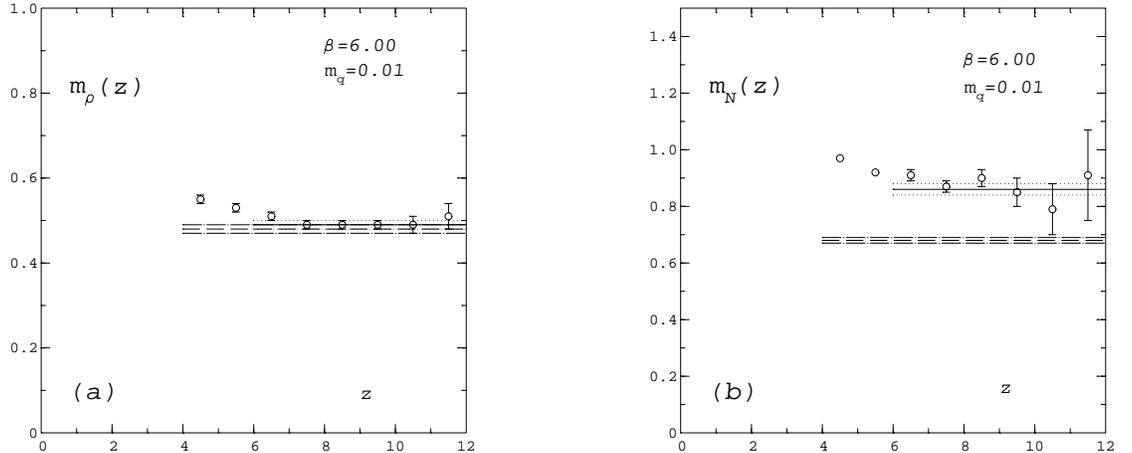

Figure 4: Figure 4(a) shows the local rho screening mass at $T = 0.92T_c$, and $m_q = 0.01$ in lattice units, from wall sources. The solid and dotted lines show the fitted value and errors, the long dashed lines the zero temperature values from [39]. Figure 4(b) shows the local nucleon screening mass at $T = 0.92T_c$, and $m_q = 0.01$ in lattice units, from wall sources. The solid and dotted lines show the fitted value and errors, the long dashed lines the zero temperature values from [39].

| Particle | $m_q = 0.01$(W) | $m_q = 0.025$(P) | $m_q = 0.035$(P) | $m_q = 0.05$(P) |
|---|---|---|---|---|
| | | $\beta = 5.90$ | | |
| $\pi$ | 0.267(2) | 0.411(3) | 0.484(2) | 0.572(2) |
| $\rho_1$ | 0.61(2) | 0.71(3) | 0.75(2) | 0.80(2) |
| $\rho_2$ | 0.55(4) | 0.63(3) | 0.70(3) | |
| $a_1$ | 0.85(10) | 1.0(1) | 1.0(1) | |
| $f_0$ | 0.67(5) | 0.77(3) | 0.83(2) | |
| $\tilde{\pi}$ | 0.35(5) | 0.52(6) | 0.60(4) | |
| N | 0.95(5) | | | |
| | | $\beta = 5.95$ | | |
| $\pi$ | 0.255(5) | 0.391(3) | | 0.550(3) |
| $\rho_1$ | 0.50(3) | 0.58(2) | | 0.72(2) |
| $\rho_2$ | 0.54(3) | 0.63(3) | | 0.70(3) |
| $a_1$ | 0.8(1) | 0.8(1) | | 1.0(1) |
| $f_0$ | 0.62(3) | 0.71(2) | | 0.83(2) |
| $\tilde{\pi}$ | 0.34(3) | 0.46(4) | | 0.62(3) |
| N | 0.95(5) | | | |
| | | $\beta = 5.975$ | | |
| $\pi$ | 0.255(2) | 0.397(2) | | 0.552(2) |
| $\rho_1$ | 0.58(6) | 0.63(3) | | 0.72(1) |
| $\rho_2$ | | 0.65(3) | | 0.70(2) |
| $a_1$ | | | | 0.85(10) |
| $f_0$ | 0.63(6) | 0.74(3) | | 0.84(1) |
| $\tilde{\pi}$ | 0.29(3) | 0.44(2) | | 0.60(1) |
| | | $\beta = 6.00$ | | |
| $\pi$ | 0.247(2) | 0.383(2) | | 0.538(2) |
| $\rho_1$ | 0.49(1) | 0.57(2) | | 0.70(1) |
| $\rho_2$ | 0.47(1) | 0.55(2) | | 0.68(1) |
| $a_1$ | 0.72(3) | 0.85(10) | | 0.97(7) |
| $f_0$ | 0.58(2) | 0.71(3) | | 0.83(1) |
| $\tilde{\pi}$ | 0.30(2) | 0.41(2) | | 0.59(1) |
| N | 0.86(3) | | | |

Table 2: Masses obtained from fits to the full propagator. W denotes wall sources, P denotes point sources. Details of the fitting procedure are discussed in section 2

# 5  Conclusions

We have analyzed the temperature dependence of various hadronic parameters. Probably the most unexpected result is that the most basic observable, the chiral condensate, does not show any significant temperature dependence up to $T = 0.92T_c$. In view of this our observation that there is no sign of temperature dependence in the pion decay constant, pion screening mass, $f_0$ screening mass or the rho screening mass may not come as a surprise. There is some, albeit inconclusive, indication of a temperature dependence for the screening mass of the nucleon.

Our calculations have been performed in the quenched approximation of QCD. The details of the temperature dependence may change in the case of QCD with two flavours, where the transition is expected to be second order rather than first order as in the pure $SU(3)$ gauge theory. However, as discussed in Section 2 we do not expect these differences to show up for temperatures less than $0.9T_c$.

# Acknowledgements

This work has been supported in part by the Stabsabteilung Internationale Beziehungen, Kernforschungszentrum Karlsruhe, a NATO research grant, contract number CRG 940451 and a DFG grant, DFG Pe-340/1. We thank the HLRZ in Jülich for the computer time used in this work.